\documentclass[12pt,a4paper]{article}
\pdfoutput=1
\usepackage{jheppub}
\usepackage{dsfont}
\usepackage{amssymb,amsmath}
\usepackage{epsfig}
\usepackage{epstopdf}
\usepackage{latexsym}
\usepackage{graphicx}
\usepackage{subfigure}
\usepackage{booktabs}
\usepackage{bbm}
\makeatletter
\def\@fpheader{\relax}
\makeatother

\def\O{{\cal O} }

\def\D{{\cal D} }
\def\T{{\cal T} }

\def\L{{\cal L} }
\def\J{{\cal J} }

\def\H{{\cal H} }

\def\S{{\cal S} }
\def\m{{M_{\text{P}}} }
\def\M{{\cal M} }
\def\t{{\bar{t}}}
\def\dcut{{%
    \setbox0\hbox{D}%
    \rlap{\hbox to \wd0{\hss ~/ \hss}}\box0}}

\def\del{\nabla}

\def\p{\partial}

\newcommand{\be}{\begin{equation}}
\newcommand{\ee}{\end{equation}}
\newcommand{\bal}{\begin{align}}
\newcommand{\eal}{\end{align}}

\def\O{{\cal O}}

\def\S{{\cal S}}

\def\L{{\cal L}}

\subheader{\begin{flushright}
UTTG-21-15\\
TCC-010-15
\end{flushright}}

\title{Membrane Paradigm, Gravitational $\Theta$-Term and Gauge/Gravity Duality}
\author{Willy Fischler$^{a,b}$, Sandipan Kundu$^{c}$}
\affiliation{$^{a}$Theory Group, Department of Physics, University of Texas, Austin, TX 78712, USA}
\affiliation{$^b$Texas Cosmology Center, University of Texas, Austin, TX 78712, USA}
\affiliation{$^c$Department of Physics, Cornell University, Ithaca, New York, 14853, USA}
\emailAdd{fischler@physics.utexas.edu}
\emailAdd{kundu@cornell.edu}

\abstract{Following the membrane paradigm, we explore the effect of the gravitational $\Theta$-term on the behavior of the stretched horizon of a black hole in $(3+1)$-dimensions. We reformulate the membrane paradigm from a quantum path-integral point of view where we interpret the macroscopic properties of the horizon as effects of integrating out the region inside the horizon. The gravitational $\Theta$-term is a total derivative, however, using our framework we show that this term affects the transport properties of the horizon. In particular,  the horizon acquires a third order parity violating, dimensionless transport coefficient which affects the way localized perturbations scramble on the horizon. Then we consider a large-N gauge theory in $(2+1)-$dimensions which is dual to an asymptotically AdS background in $(3+1)-$dimensional spacetime  to show that the $\Theta$-term induces a non-trivial contact term in the energy-momentum tensor of the dual theory. As a consequence, the dual gauge theory in the presence of the $\Theta$-term acquires the same third order parity violating transport coefficient.}

\begin{document}

\maketitle
\flushbottom

\section{Introduction}
Black holes are not only fascinating but they also provide us with a natural laboratory to perform thought experiments to understand quantum gravity. String theory, Matrix Theory \cite{Banks:1996vh}, and the AdS/CFT correspondence \cite{Maldacena:1997re}, which are the only models of quantum gravity over which we have mathematical control, have provided us with some insight into different aspects of quantum gravity, \emph{e.g.} the Bekenstein-Hawking entropy formula for a large class of black holes. They also strongly indicate that black hole evolution as seen by an external observer is unitary. However, none of these models give us a comprehensive microscopic description of the physics of black holes.

Historically the membrane paradigm \cite{Price:1986yy,Thorne:1986iy} has also been successful at providing us with a powerful framework to study macroscopic properties of black hole horizons. In astrophysics the membrane paradigm has been used extensively as an efficient computational tool to study phenomena in the vicinity of black holes (see \cite{Thorne:1986iy, Anninos:1994gp,Masso:1998fi,Komissarov:2004ms, Penna:2013rga, Penna:2015qta} and references therein). The membrane paradigm has also been able to provide crucial hints about details of the microscopic physics of horizons. In particular, the membrane paradigm predicts that black hole horizons are the {\it fastest scramblers} in nature. Fast-scrambling strongly indicates that the microscopic description of scrambling of information on static horizons must involve non-local degrees of freedom \cite{Sekino:2008he,Susskind:2011ap}. In this paper, we will try to understand the membrane paradigm from a quantum path-integral point of view. We will interpret the macroscopic properties of the horizon as effects of integrating out the region inside the horizon.  The semi-classical approximation of this path-integral approach is equivalent to the action formulation \cite{Parikh:1997ma} of the conventional membrane paradigm.

We are mainly interested in figuring out how total derivative terms can affect the macroscopic properties of black hole horizons. Total derivative terms do not affect the classical equations of motion and hence do not contribute even in perturbative quantum field theory. However,  it is well known that total derivative terms can have physical effects, \emph{e.g.} Lorentz and gauge invariance of Quantum chromodynamics (QCD) allow for a CP-violating topological $\theta_{QCD}$ term which contributes to the electric dipole moment of  neutrons \cite{Crewther:1979pi}. Similarly, the electrodynamics $\theta$-term is also a total derivative, therefore, does not contribute for perturbative quantum electrodynamics (QED). However, in the presence of the electrodynamics $\theta$-angle a black hole horizon behaves as a Hall conductor, for an observer hovering outside \cite{Fischler:2015cma}. As a consequence, the electrodynamics $\theta$-angle affects the way localized perturbations, created on the stretched horizon by dropping a charged particle, {\it fast scramble} on the horizon \cite{Fischler:2015cma}. Later it was also shown that in-falling electric charges produce a non-trivial Berry phase in the QED wave function which can have physical effects in the early universe \cite{Banks:2015xsa}.

Another example comes from gravity  in $(3+1)-$dimensions, where the topological Gauss-Bonnet term contributes a correction term to the entropy of a black hole which is proportional to the Euler number of the horizon. One can show that this correction term violates the second law of black hole thermodynamics and hence should be zero \cite{Jacobson:1993xs,Olea:2005gb, Liko:2007vi, Sarkar:2010xp}. In $(3+1)-$dimensions, there exists another total derivative term, a parity violating gravitational $\Theta$-term
\begin{equation}\nonumber
 S_\Theta=\frac{\Theta}{8}\int d^4x\ \epsilon^{\mu \nu \alpha \beta}R^\tau_{~\sigma \mu \nu} R^{\sigma}_{~\tau \alpha \beta}\ .
\end{equation}
In this paper, we explore the effect of this $\Theta$-term on black hole horizons.  The membrane paradigm tells us that for an outside observer a black hole horizon effectively behaves like a viscous Newtonian fluid. Using our framework, we will show that  the gravitational $\Theta$-term affects the transport properties of the horizon fluid, in particular,  the horizon fluid acquires a third order parity violating, dimensionless transport coefficient, which we will call $\vartheta$. This indicates that the $\Theta$-term will affect the way perturbations scramble on the horizon. Specifically, we can perform a thought experiment, in which an outside observer drops a massive particle onto the black hole and watches how the perturbation scrambles on the black hole horizon. We will argue that the gravitational $\Theta$-term, similar to the electrodynamics $\theta$-term, will also introduce vortices without changing the scrambling time. This strongly suggests that in a sensible theory of quantum gravity the $\Theta$-term will play an important role, a claim that we will show is also supported by the AdS/CFT correspondence.

The membrane paradigm has become even more relevant with the emergence of {\it holography} \cite{Banks:1996vh, Maldacena:1997re}, a remarkable idea that connects two cornerstones of theoretical physics: quantum gravity and gauge theory. The AdS/CFT correspondence \cite{Maldacena:1997re}, which is a concrete realization of this idea of holography, has successfully provided us with theoretical control over a large class of strongly interacting field theories \cite{Maldacena:1997re, Witten:1998qj, Gubser:1998bc, Aharony:1999ti}. This duality enables us to compute observables of certain large-N gauge theories in $d$-dimensions by performing some classical gravity calculations in $(d+1)$-dimensions. Gravity duals of these field theories at finite temperature contain black holes with horizons.  It has been shown that there is some connection between the low frequency limit of linear response of a strongly coupled quantum field theory and the membrane paradigm fluid on the black hole horizon of the dual gravity theory \cite{Kovtun:2003wp,Kovtun:2004de,Son:2007vk,Iqbal:2008by,Bredberg:2010ky}. In this paper, we will consider a large-N gauge theory in $(2+1)-$dimensions which is dual to a gravity theory in $(3+1)-$dimensions with the gravitational $\Theta$-term and figure out the effect of the parity violating $\Theta$-term on the dual field theory. A reasonable guess is that the boundary theory, similar to the membrane paradigm fluid, will acquire the same third order parity violating transport coefficient $\vartheta$.  We will confirm this guess by performing an explicit computation. 

It was argued in \cite{Closset:2012vp} that the two-point function of the energy-momentum tensor in a $(2+1)-$dimensional conformal field theory can have a non-trivial contact term
\be\nonumber
\langle T_{ij}(x)T_{mn}(0)\rangle=-i\frac{\kappa_g}{192\pi} \left[\left(\varepsilon_{iml}\partial^l\left(\partial_j \partial_n-\partial^2 \delta_{jn} \right)+(i\leftrightarrow j)\right)+(m\leftrightarrow n)\right]\delta^3(x)\ .
\ee
It is possible to shift $\kappa_g$ by an integer by adding a gravitational Chern-Simons counterterm to the UV-Lagrangian and hence the integer part of $\kappa_g$ is scheme-dependent. On the other hand, the fractional part $\kappa_g$ mod $1$  does not depend on the short distance physics and hence it is a meaningful physical observable in $(2+1)-$dimensional conformal field theory \cite{Closset:2012vp}. We will argue that a gravity theory in AdS$_{(3+1)}$  with the gravitational $\Theta$-term is dual to a conformal field theory with  non-vanishing $\kappa_g$, in particular
\be\nonumber
\frac{\kappa_g}{96\pi}=\Theta 
\ee
which also suggests that only a fractional part of the $\Theta$-term is a well-defined observable.\footnote{This also suggests that $\Theta$, in our normalization is not an angle. However, one can work in the normalization in which the gravitational $\Theta$-term takes the form
\begin{equation}\nonumber
 S_\Theta=\frac{\Theta}{1536 \pi^2}\int d^4x\ \epsilon^{\mu \nu \alpha \beta}R^\tau_{~\sigma \mu \nu} R^{\sigma}_{~\tau \alpha \beta}\ .
\end{equation}
In that case, adding an integer to $\kappa_g$ changes the new $\Theta$ by an integer times $2\pi$ and hence in the above normalization  $\Theta$ is an angle.
 }

The contact term $\kappa_g$ is also related to the transport coefficient $\vartheta$, which to our knowledge has never been studied before. It is a parity violating third order\footnote{Little is known about third-order transport coefficients in any dimensions. Very recently, third order hydrodynamics for neutral fluids has been studied in $(3+1)$-dimensions \cite{Grozdanov:2015kqa}.}  transport coefficient  in $(2+1)-$dimensions and hence forbidden in a parity-invariant theory. Under a small metric perturbation $\gamma_{AB}$ around flat Minkwoski metric, it contributes to the energy-momentum tensor in the following way:
\begin{align}
T_{11}=-T_{22}=-\vartheta \frac{\partial^3 \gamma_{12}}{\partial t^3}\ ,\qquad
T_{12}=T_{21}=\frac{\vartheta}{2 }\left( \frac{\partial^3 \gamma_{11}}{\partial t^3}- \frac{\partial^3 \gamma_{22}}{\partial t^3}\right) \nonumber\ 
\end{align}
and hence $\vartheta$ contributes to the retarded Green's function of the energy-momentum tensor in order $\omega^3$:
\begin{align}
G^R_{12,11-22}(\omega,\vec{k}\rightarrow 0)=-2i \vartheta \omega^3\ .\nonumber
\end{align}
$\vartheta$ is dimensionless and it does not affect the trace of the energy-momentum tensor. In $(2+1)-$dimensional hydrodynamics, the {\it Hall viscosity} is another parity violating effect that appears in the first order in derivative expansion. The Hall viscosity has been studied extensively for both relativistic \cite{Saremi:2011ab} and non-relativistic systems \cite{Avron:1995fg, Avron1997, Read:2010epa}. We believe that $\vartheta$ is a third order cousin of Hall viscosity and hence it is also an example of Berry-like transport \cite{Haehl:2014zda}. We will show that for a holographic theory dual to asymptotically AdS spacetime in $(3+1)-$dimensions: $\vartheta=\Theta=\kappa_g/96\pi$. We will also speculate on the possible covariant structure of the $\vartheta$ contribution to the energy-momentum tensor. 

The rest of the paper is organized as follows. We start with a discussion of the membrane paradigm in section \ref{mp}. In section \ref{mpg}, we review the membrane paradigm for the Einstein gravity. Then in section \ref{theta}, we introduce gravitational $\Theta$-term and discuss its effect on the stretched horizon. In section \ref{adscft}, we discuss the effect of the $\Theta$-term in the context of the AdS/CFT correspondence and make some comments on the $\vartheta$-transport in section \ref{section_kubo}. Finally, we conclude in section \ref{conclude}. Some technical details have been relegated to appendices \ref{nh} and \ref{transport}. For readers only interested in the effect of the $\Theta$-term in the context of the AdS/CFT correspondence, it is sufficient to read sections  \ref{adscft} and \ref{section_kubo}.

\section{Integrating out inside: membrane paradigm}\label{mp}

The membrane paradigm provides a simple formalism to study macroscopic properties of horizons by replacing the true mathematical horizon by a {\it stretched horizon}, an effective time-like membrane located roughly one Planck length away from the true horizon. Finiteness of the black hole entropy suggests that between the actual black hole horizon and the stretched horizon, the effective number of degrees of freedom should be vanishingly small. So, it is more natural as well as convenient to replace the true mathematical horizon by a stretched horizon.

Predictions of the membrane paradigm are generally considered to be robust since they depend on some very general assumptions: 

\begin{itemize}
\item{ The effective number of degrees of freedom between the actual black hole horizon and the stretched horizon are vanishingly small.} 
\item { Physics outside the black hole, classically must not be affected by the dynamics inside the black hole. }
\end{itemize}
In this section, we will try to reformulate the membrane paradigm from a quantum path-integral point of view. Our goal is to interpret the macroscopic properties of the stretched horizon as effects of integrating out the region inside the stretched horizon.\footnote{A discussion on integrating out geometry in the context of the AdS/CFT correspondence can be found in \cite{Faulkner:2010jy}.} We will not attempt to derive an effective action of the membrane, rather we will work in the semi-classical approximation where a lot can be learnt even without knowing the exact membrane effective action. However, our approach is somewhat similar to the approach of \cite{CaronHuot:2011dr} and it should be possible to derive a  
quantum mechanical version of the membrane paradigm from our approach by following \cite{CaronHuot:2011dr}.

\subsection{Fields in the black hole background}
Let us first consider some fields in a black hole background.  We will assume that the back-reactions of the fields to the background is negligible. The action is given by
\begin{equation}\label{action}
S[\phi_I]=\int d^{d+1}x \sqrt{g}\L(\phi_I, \del_\mu \phi_I)\ ,
\end{equation}
where $\phi_I$ with $I=1,2,...$ stands for any fields.  It is necessary to impose some boundary conditions on the fields $\phi_I$ in order to obtain equations of motion by varying this action. We will impose  Dirichlet boundary conditions $\delta \phi_I =0$ at the boundary of space-time. The stretched horizon $\M$ divides the whole space-time in regions: 
\begin{align}
A:&~\qquad \text{outside the membrane}~ \M \ ,\nonumber\\
B:&~\qquad \text{inside the membrane}~ \M\ .\nonumber 
\end{align}
The total quantum field theory partition function for fields $\phi_I$ is given by,
\be
Z=\int \D[\phi_I]e^{i S[\phi_I]}\ .
\ee

Now imagine an observer $\O$ who is hovering outside the horizon of a black hole. Observer $\O$ has access only to the region outside the black hole. We can write down the above partition function in the following way
\be\label{pi1}
Z=\int \D[\phi^B_I]\D[\phi^\M_I]\D[\phi^A_I] e^{i S_B[\phi_I]}e^{i S_A[\phi_I]}\ .
\ee
Where, we have written $S[\phi_I]=S_A[\phi_I]+S_B[\phi_I]$. In the path integral, we have decomposed every field $\phi$ such that $\phi^B$ is the the field inside the stretched horizon $\M$, $\phi^A$ is the the field outside the stretched horizon $\M$ and $\phi^\M$ is the the field on the stretched horizon $\M$. Observer $\O$ has access only to the region outside the black hole and we want to find out some effective action $S_\O$ for the observer $\O$. To that goal, we first fix the values of all the fields $\phi_I$ with $I=1,2,...$ on $\M$. Then in principle we can perform the path integral $\int \D[\phi^B_I]e^{i S_B[\phi_I]}\equiv z(\phi^\M_I)$. After performing the path integral over $\phi^B_I$, partition function (\ref{pi1}) becomes,
\be\label{pi2}
Z=\int \D[\phi^\M_I]\D[\phi^A_I]z(\phi^\M_I) e^{i S_A[\phi_I]}\ .
\ee
This partition function now depends only on quantities defined on or outside of the stretched horizon $\M$. The non-trivial function $z(\phi^\M_I)$ contains information about the inside. In practice, it is not possible to find $z(\phi^\M_I)$ because that requires detailed knowledge of the physics inside the black hole. Our goal is not to compute $z(\phi^\M_I)$ exactly, but to extract some information about $z(\phi^\M_I)$ by demanding that the classical physics outside the black hole horizon must not be affected by the dynamics inside the black hole.

Let us now, re-write the partition function (\ref{pi2}), in the following way:
\be\label{pi3}
Z_\O=\int_A \D[\phi_I]e^{i (S_A[\phi_I]+S_{surf}[\phi_I])}\ ,
\ee
where, now fields $\phi_I$'s are defined only on or outside of the stretched horizon $\M$ and $z(\phi^\M_I)\equiv \exp(iS_{surf}[\phi_I])$. Observer $\O$ has access only to the region outside the black hole and physics he observes, classically must not be affected by the dynamics inside the black hole. That means observer $\O$ should be able to obtain the correct classical equations of motion by varying action $S_\O$ which is restricted only to the space-time outside the black hole:
\be
S_\O=S_A[\phi_I]+S_{surf}[\phi_I]\ .
\ee
Note that $S_{surf}[\phi_I] \neq 0$ because the boundary terms generated on $\M$ from $S_A$ are in general non-vanishing. Surface terms $S_{surf}[\phi_I]$ obtained by integrating over fields inside the stretched horizon must exactly cancel all these boundary terms. The fact that the correct classical equations of motion can be obtained by varying only $S_\O=S_A[\phi_I]+S_{surf}[\phi_I]$, gives us certain information about $S_{surf}[\phi_I]$. For the observer $\O$, the action  $S_\O$ for fields $\phi_I$  now have sources residing on the stretched horizon
\begin{align}
S_\O=\int_A d^{d+1}x \sqrt{-g}\L(\phi_I, \del_\mu \phi_I)+\sum_I \int_\M d^{d}x \sqrt{-h} \ \J_\M^I \phi_I
\end{align}
where, $h$ is the determinant of the induced metric on the stretched horizon $\M$ and we have written
\be
S_{surf}[\phi_I]=\sum_I \int_\M d^{d}x \sqrt{-h} \ \J_\M^I \phi_I\ .
\ee
It is important to note that one should interpret $\J_\M^I$ as external sources such that $\frac{\delta \J_\M^I}{\delta \phi_J}=0$. Now demanding that we obtain correct classical equations of motion for fields $\phi_I$ by varying $S_\O$, we obtain
\be\label{source}
\J_\M^I= \left[n_\mu \frac{\partial \L}{\partial \left(\del_\mu \phi_I \right)}\right]_\M
\ee
where, $n_\mu$ is the outward pointing normal vector to the time-like stretched horizon $\M$ with $n_\mu n^\mu =1$. The observer $\O$ can actually perform real experiments on the stretched horizon $\M$ to measure the sources $\J_\M^I$. 


\subsection{Example: Electrodynamics $\theta$-term and stretched horizon}
Several examples of the action principle of the membrane paradigm can be found in \cite{Parikh:1997ma}. However, we will focus on a particular example studied in \cite{Fischler:2015cma} which shows that total derivative terms can have important physical effects. The action for electromagnetic fields with a $\theta$-term in curved space-time in $(3+1)-$dimensions is
\begin{align}\label{actioncs}
S=\int \sqrt{g}d^4x \left[- \frac{1}{4}F_{\mu\nu}F^{\mu \nu}+j_\mu A^\mu \right]+\frac{\theta}{8} \int d^4x \epsilon^{\alpha \beta \mu \nu}F_{\alpha \beta}F_{\mu \nu}
\end{align}
where,  $F_{\mu\nu}=\p_\mu A_\nu-\p_\nu A_\mu $. Current  $j^\mu$ is conserved, i.e., $\del_\mu j^\mu=0$.\footnote{Our convention of the metric is that the Minkowski metric has signature $(-+++)$.} The electrodynamics $\theta$-term does not affect the classical equations of motion because it is a total derivative and hence the equations of motion obtained from the action (\ref{actioncs}) is
\begin{align}\label{meq}
\del_\mu F^{\mu\nu}=-j^\nu\ .
\end{align}
Field strength tensor $F_{\mu\nu}$ also obeys $\p_{[\mu}F_{\nu \lambda]}=0$. Let us now write 
\be
\frac{\theta}{8}\epsilon^{\alpha \beta \mu \nu}F_{\alpha \beta}F_{\mu \nu}=\frac{\theta}{4}\sqrt{g}F_{\mu \nu}*F^{\mu\nu}\ .
\ee
Our convention of the Levi-Civita tensor density $\epsilon^{\alpha \beta \mu \nu}$ is the following: $\epsilon^{0123}=1$, $\epsilon_{0123}=-g$. We will also assume that the conserved current $j^\mu$ is contained inside the membrane $\M$ and hence the observer $\O$ who has access only to the region outside the stretched horizon does not see the current $j^\mu$. However, the observer will see a surface current $\J_\M^\mu$ on the membrane. Let us start with the action for the observer $\O$
\begin{align}\label{action1}
S_\O=\int_A \sqrt{g}d^4x \left[- \frac{1}{4}F_{\mu\nu}F^{\mu \nu}+\frac{\theta}{4}F_{\mu \nu}*F^{\mu\nu} \right]+\int_\M \sqrt{-h}  d^{3}x \J_{\M;\mu} A^\mu \ .
\end{align}
The action is invariant under the gauge transformation: $A_\mu \rightarrow A_\mu +\p_\mu \lambda $ only if $\J_{\M;\mu}n^\mu=0$, where $n_\mu$ is the outgoing unit normal vector on $\M$. In order for the observer $\O$ to recover the vacuum Maxwell's equations, outside the horizon, the boundary terms on $\M$ should cancel out and from equation (\ref{source}) we obtain
\be\label{hcurrent}
 \J_\M^\mu=\left(n_\nu F^{\mu\nu}-\theta\ n_\nu *F^{\mu\nu}\right)|_\M\ .
\ee
Note that $\J_{\M;\mu}n^\mu=0$ and hence the action (\ref{action1}) is invariant under gauge transformations. 

The electrodynamics $\theta$-term is a total derivative, therefore, it does not contribute to perturbative quantum electrodynamics (QED), which indicates that the effects of the $\theta$ term in QED, if any, are non-perturbative and hence difficult to detect. But by coupling this theory to gravity  one finds that the $\theta$-term can affect the electrical properties of black hole horizons,  in particular, black hole horizons behave as {\it Hall conductors} \cite{Fischler:2015cma}. This strongly suggests that in a sensible quantum theory of black holes, a total derivative term like electrodynamics $\theta$-term can play important role. 

The AdS/CFT correspondence which is one of the few models of quantum gravity which is well understood also supports the above claim  \cite{Fischler:2015cma}. In particular, let us consider U(1) gauge field in AdS-Schwarzschild in $(3+1)-$dimensions with the action
\begin{align}
S=\int d^4x \left[- \frac{\sqrt{g}}{4g_{3+1}^2}F_{\mu\nu}F^{\mu \nu} +\frac{\theta}{8}  \epsilon^{\alpha \beta \mu \nu}F_{\alpha \beta}F_{\mu \nu}\right]\ .
\end{align}
The U(1) gauge field in the bulk is dual to a conserved current $j^i$ in the boundary theory. DC conductivities are given by $\sigma^{AB}=i \lim_{\omega \rightarrow 0} \lim_{\mathbf{k} \rightarrow 0} G_R^{AB}(\mathbf{k})/ \omega$, where, $G_R^{AB}(\mathbf{k})$ is the retarded Green function of boundary current $j$ and indices $A,B=1,2$ run over the spatial directions of the boundary theory. The DC conductivity $\sigma^{AB}$ of the strongly coupled $(2+1)-$dimensional dual theory is given by \cite{Iqbal:2008by}
\be
\sigma^{11}=\sigma^{22}=\frac{1}{g_{3+1}^2}\ , \qquad \sigma^{21}=-\sigma^{12}=\theta\ 
\ee
and hence in the presence of the $\theta$-term, the boundary theory has nonzero Hall conductivity. 
\section{Membrane paradigm and gravity}\label{mpg}
First let us consider the Einstein-Hilbert action for gravity
\begin{equation}\label{EH}
 S_{EH}=\frac{\m^2}{2}\int d^4x\sqrt{-g}R \ .
\end{equation}
A theory of gravity contains higher derivatives in the action and hence we should be more careful. In order to obtain equations of motion by performing a variation of this action, it is not sufficient to impose Dirichlet boundary conditions $\delta g^{\mu\nu}=0$ on  the outer boundary of space-time $\Sigma$. With the Dirichlet boundary conditions on $\Sigma$, variation of the Einstein-Hilbert action gives rise to non-vanishing boundary terms. One solution is to further impose $\del_\rho\delta g^{\mu\nu}=0$ on $\Sigma$. However, it is more useful to modify the action (\ref{EH}) by adding the standard Gibbons-Hawking-York boundary term to make the action  consistent with the Dirichlet boundary condition on the boundary $\Sigma$
\begin{equation}\label{ehghy}
S=S_{EH}+S_{GHY}(\Sigma)=\frac{\m^2}{2}\int d^4x\sqrt{-g}R +\m^2\int_\Sigma d^3 x\sqrt{|h|}K \ ,
\end{equation}
where, $K$ is the trace of the extrinsic curvature and $h$ is the determinant of the induced metric on $\Sigma$.

Let us again imagine a stretched horizon $\M$ that divides the whole space-time in two regions
\begin{align}
A:&~\qquad \text{outside the membrane}~ \M \ ,\nonumber\\
B:&~\qquad \text{inside the membrane}~ \M\ \nonumber 
\end{align}
and an observer $\O$ who has access only to the region outside the stretched horizon. The total quantum partition function can in principle be written in the following way
\be\label{g1}
Z=\int \D[g_{\mu\nu}^B]\D[g_{\mu\nu}^\M]\D[g_{\mu\nu}^A] e^{i S_B}e^{i S_A}\ .
\ee
Where, we have written the total action $S=S_A+S_B$. Observer $\O$ has access only to the region outside the black hole and we want to find out some effective action $S_\O$ for the observer $\O$. Since, $\delta S_A \neq \int_\M (...)\delta g_{\mu\nu}$, one must divide the total action (\ref{ehghy}) in the following way:
\be\label{insideout}
S=\left(S_A-S_{GHY}(\M)\right)+\left(S_B+S_{GHY}(\M)\right)\ 
\ee
because otherwise the action is non-differentiable with respect to the metric on $\M$. The path integral (\ref{g1}), can now be written as
 \be\label{g2}
Z=\int_{A+\M} \D[g_{\mu\nu}^\M]\D[g_{\mu\nu}^A] e^{i \left(S_A-S_{GHY}(\M)\right)}\int_B\D[g_{\mu\nu}^B]e^{i \left(S_B+S_{GHY}(\M)\right)}\ .
\ee
Now we first fix the metric on $\M$ and then in principle we can perform the path integral on $B$. Performing this path integral exactly is an impossible task without the detailed knowledge of the physics inside the black hole. But fortunately we do not need to perform the path integral, instead we write
\be
\int_B\D[g_{\mu\nu}^B]e^{i \left(S_B+S_{GHY}(\M)\right)}=e^{i S_{surf}(\M)}\ .
\ee
Hence, the partition function (\ref{g2}) becomes,
 \be\label{g3}
Z_\O=\int_{A+\M} \D[g_{\mu\nu}]e^{i \left(S_A-S_{GHY}(\M)+S_{surf}(\M)\right)}\ ,
\ee
where, now the path integral is defined only on or outside of the stretched horizon $\M$. Observer $\O$ has access only to the region outside the black hole and physics he observes, classically must not be affected by the dynamics inside the black hole. That means observer $\O$ should be able to obtain the correct classical equations of motion by varying action $S_\O$ which is restricted only to the space-time outside the black hole, where,
\begin{align}\label{ehgravity}
 S_\O =& S_{EH}+S_{GHY}(\Sigma)-S_{GHY}(\M)+S_{surf}(\M)\\
 =&\int_A d^4x[\frac{\m^2}{2}\sqrt{-g}R]+\m^2\int_{\Sigma} d^3 x\sqrt{|h|}K-\m^2\int_{\M} d^3 x\sqrt{|h|}K+S_{surf}(\M)\nonumber\ .
\end{align}
Note that the sign of the $S_{GHY}(\M)$ term is negative because we choose outward pointing normal vector to be positive. Possibly, one can interpret the above action in the following way. The observer $\O$ has two boundaries: outer space-time boundary $\Sigma$ and another boundary because of the membrane $\M$ and hence he needs the Gibbons-Hawking-York boundary term for both $\Sigma$ and  $\M$. The boundary conditions are fixed only at the outer boundary $\Sigma$ and the surface term $S_{surf}(\M)$ is necessary to obtain the correct classical equations of motion.

It is important to note that the division of the action (\ref{insideout}) is not unique. We can always add some intrinsic terms,
\be
S=\left(S_A-S_{GHY}(\M)+S_i(\M)\right)+\left(S_B+S_{GHY}(\M)-S_i(\M)\right)\ ,
\ee  
where,
\be
S_i(\M)=\int_\M d^3x\sqrt{h} F(h_{ij})\ .
\ee
$F(h_{ij})$ can be any scalar intrinsic to $\M$. Since, it's an intrinsic term, now
\be
\int_B\D[g_{\mu\nu}^B]e^{i \left(S_B+S_{GHY}(\M)-S_i(\M)\right)}=e^{i( S_{surf}(\M)-S_i(\M))}\ 
\ee  
and hence $S_\O$ does not depend on $S_i(\M)$. 

In the rest of the section, we will mainly review some known results of the membrane paradigm. Experts can skip the rest of this section and move on to section \ref{theta}.
\subsection{Energy momentum tensor on the stretched horizon}

Let us now consider variations of $S_{EH}$ and $S_{GHY}(\Sigma)$ with Dirichlet boundary conditions on $\Sigma$
\begin{align}
\delta (S_{EH}+S_{GHY}(\Sigma))=&\frac{\m^2}{2}\int_A d^4x\left[\delta \left(\sqrt{-g}g^{\mu \nu}\right)R_{\mu \nu}+\sqrt{-g}g^{\mu \nu} \delta R_{\mu \nu} \right]\nonumber\\
=&\frac{\m^2}{2}\int_A d^4x\sqrt{-g}\left[G_{\mu \nu} \delta g^{\mu \nu} +\sqrt{-g}g^{\mu \nu} \delta R_{\mu \nu} \right]\ ,\nonumber\\
\end{align}
where $G_{\mu \nu}=R_{\mu \nu}-\frac{1}{2}g_{\mu \nu} R$ is the Einstein tensor. The variation of the Ricci tensor is given by,
\begin{equation}
g^{\mu \nu} \delta R_{\mu \nu}=-\nabla_\mu \left[\nabla_\nu \delta g^{\mu \nu} -  g_{\alpha \nu} g^{\mu \beta}\nabla_\beta \delta g^{\alpha \nu} \right]\ .
\end{equation}
Therefore,\footnote{Let us recall that the Gauss' theorem in curved space-time is given by
\begin{equation}
\int_V \sqrt{-g}d^4 x \nabla_\mu A^\mu=\oint_{\partial V}\sqrt{|h|}d^3x n_\mu A^\mu\ ,
\end{equation}
where $h$ is the induced metric on the surface $\partial V$ and $n_\mu$ is the normal vector with $n_\mu n^\mu =1$ (assuming the surface is timelike).
}

\begin{align}\label{eqn1}
\int_A d^4x \sqrt{-g}g^{\mu \nu} \delta R_{\mu \nu}&=\int_\M d^3x \sqrt{|h|}n_\mu g_{\alpha \nu}\left[\nabla^\alpha \delta g^{\mu \nu} -  \nabla^\mu \delta g^{\alpha \nu} \right]\nonumber\\
&=\int_\M d^3x \sqrt{|h|}n_\mu h_{\alpha \nu}\left[\nabla^\alpha \delta g^{\mu \nu} -  \nabla^\mu \delta g^{\alpha \nu} \right]\ .
\end{align}
$n_\mu$ is the normal vector to the time-like surface $\M$ with $n_\mu n^\mu =1$. Note that there is a negative sign because $n^\mu$ is the outward pointing normal vector. $h$ is the determinant of the induced 3-metric on the membrane $\M$. The induced 3-metric has been written as a 4-metric 
\be
h^{\mu \nu}=g^{\mu \nu}-n^\mu n^\nu
\ee 
that projects from the 4-dimensional space-time to the 3-dimensional membrane $\M$. The membrane extrinsic curvature is defined as
\be\label{k}
K_{\mu \nu}\equiv h_{\mu}^{~\alpha}h_{\nu}^{~\beta}\nabla_\alpha n_\beta\ .
\ee
Let us first note some of the properties of the extrinsic curvature tensor (\ref{k}). One can easily check that
\be
K_{\mu\nu}n^\mu=K_{\mu\nu}n^\nu=0\ .
\ee
One can also check that $K_{\mu\nu}=K_{\nu\mu}$.\footnote{Note that the vector $n^\mu$ is orthogonal to $\M$ and hence it obeys the hypersurface orthogonality condition: $n_{[\mu}\nabla_\nu n_{\lambda]}=0$. }

Let us now compute the variation of $S_{GHY}(\M)$
\begin{align}
\delta S_{GHY}(\M)=&\m^2 \int_{\M} d^3 x ~\delta(\sqrt{|h|}h^{\mu \nu} \nabla_\mu n_\nu )\nonumber\\
=& \m^2 \int_{\M} \sqrt{|h|}d^3 x \left[-\frac{1}{2}h_{\mu \nu}K\delta h^{\mu \nu} +\frac{1}{2}K_{\mu \nu}\delta h^{\mu \nu}+\nabla_\mu c^\mu\right]\nonumber\\
&~~~~~+\frac{1}{2}\m^2\int_\M d^3x \sqrt{|h|}n_\mu h_{\alpha \nu}\left[\nabla^\alpha \delta g^{\mu \nu} -  \nabla^\mu \delta g^{\alpha \nu} \right]\ ,
\end{align}
where,
\begin{equation}
c_\mu=\delta n_\mu-\frac{1}{2}n^\nu \delta g_{\mu\nu}\ .
\end{equation}
One can easily check that $n^\mu c_\mu=0$ and as a consequence
\begin{equation}
\nabla_\mu c^\mu|_\M=D^{(3)}_i c^i |_\M\ ,
\end{equation}
where $D^{(3)}_i$ is the covariant derivative in terms of the 3-dimensional induced metric $h_{ij}$. Therefore, the term $\nabla_\mu c^\mu$ in $\delta S_{GHY}(\M)$ is a total derivative and hence does not contribute. So, finally we obtain,
\begin{align}
\delta (S_{EH}+S_{GHY}(\Sigma)-\delta S_{GHY}(\M))&=\frac{\m^2}{2}\int_A d^4x\sqrt{-g}G_{\mu \nu} \delta g^{\mu \nu}\nonumber\\
 &+\frac{\m^2}{2} \int_{\M} \sqrt{|h|}d^3 x \left[h_{\mu \nu}K -K_{\mu \nu}\right]\delta h^{\mu \nu}\ .
\end{align}
Therefore all the terms in the second line should be cancelled by $\delta S_{surf}$. The membrane energy-momentum tensor $\T_{\mu \nu}$ is defined in the standard way
\begin{equation}
\delta S_{surf}(\M)=-\delta (S_{EH}+S_{GHY}(\Sigma)-\delta S_{GHY}(\M))=-\frac{1}{2} \int_{\M} \sqrt{|h|}d^3 x \T_{\mu \nu}\delta h^{\mu \nu}\ .
\end{equation}
Therefore, finally we obtain
\begin{equation}\label{membrane_em}
 \T_{\mu \nu}=\m^2\left[h_{\mu \nu}K -K_{\mu \nu}\right]|_\M\ .
\end{equation}
It is important to note that $\T_{\mu \nu} n^\mu=0$. This is the famous result that was originally obtained in \cite{Price:1986yy} from equations of motion and later in  \cite{Parikh:1997ma} from action formulation of membrane paradigm.

\subsection{Near horizon geometry}
Now we will use equation (\ref{membrane_em}) to review the claim that the stretched horizon behaves as a viscous Newtonian fluid. For that we only need to know the near horizon geometry.   Let us denote the 3-dimensional absolute event horizon by $\H$. We define a well-behaved time coordinate $\t$ on the horizon as well as spatial coordinates $x^A$ with $A=1,2$. Therefore the event horizon $\H$ has coordinates $x^i\equiv \left(\t, x^A\right)$ with a metric $h_{ij}$. There exists a unique null generator $l^\mu$ through each point on the horizon (in the absence of horizon caustics). We can always choose coordinates $x^i\equiv \left(\t, x^A\right)$ such that
\begin{equation}
l=\frac{\partial}{\partial \t}\ .
\end{equation}
Where we have chosen the spatial coordinates $x^A$ such that they comove with the horizon generator $l$ and $h_{0A}=0$. Note that
\begin{equation}\label{gh}
l^\mu \nabla_\mu l^\nu= g_H l^\nu \ ,
\end{equation}
where $g_H$ is the surface gravity of the horizon $\H$. We only restrict to the simpler case: $g_H(\t,x) =$constant on the horizon $\H$.  For the case of constant and non-zero $g_H$,  the near horizon geometry has the form (see appendix \ref{nh} for details)
\begin{equation}\label{nhmetric}
ds^2=-r^2~d\t^2+ \frac{2r}{g_H}d\t dr+\gamma_{AB}\left(dx^A-\frac{\Omega^A(\t, x)}{g_H} r^2 d\t \right)\left(dx^B-\frac{\Omega^B(\t, x)}{g_H} r^2 d\t \right)+\O (r^4)
\end{equation}
where the horizon is at $r=0$ and the induced spatial metric on a constant $\t$ hypersurface is $\gamma_{AB}\equiv\gamma_{AB}(\t,x,r)$. Let us also note that in comoving coordinates
\begin{equation}
\frac{\partial \gamma_{AB}}{\partial \t}=2 \sigma_{AB}^H+\theta_H \gamma_{AB}\ ,
\end{equation}
where, horizon expansion $\theta_H$ and shear $\sigma_{AB}^H$ are defined as 
\begin{align}
&\sigma_{AB}^H=\theta_{AB}-\frac{1}{2}\gamma_{AB}\theta_H\ , \\
&\theta_H= \gamma^{AB}\theta_{AB}=\frac{\partial}{\partial \t}\ln \sqrt{\gamma}\ .
\end{align}
$\theta_{AB}$ is the 2-dimensional covariant derivative (with metric $\gamma_{AB}$) of the horizon null generator
\begin{equation}
\theta_{AB}=D^{(2)}_A l_B\ .
\end{equation}¥

We now have to choose a set of FIDO's and a universal time. $\t$ is not a good choice for the universal time because surfaces of constant $\t$ are null everywhere. However, there is a preferred choice for the universal time $t$ \cite{Price:1986yy}
\begin{equation}
t=\t-\frac{1}{g_H}\ln r\ .
\end{equation}
FIDO's are chosen such that their world lines are orthogonal to constant $t$ surfaces and hence velocity 4-vector can be written as
\begin{equation}
U=-rdt=-rd\t+\frac{dr}{g_H}\ .
\end{equation}
Therefore,
\begin{equation}
U^r=0\ , \qquad U^\t=\frac{1}{r}\ , \qquad U^A=-\frac{\Omega^A r}{g_H}\ .
\end{equation}
Note that in the limit $r \rightarrow 0$, these world lines approach the null generator of the horizon.

Let us now replace the actual horizon by a stretched horizon at $r=\epsilon$. In the limit $\epsilon \rightarrow 0$, using the near horizon metric (\ref{nhmetric}), components of the extrinsic curvature (\ref{k}) are given by (up to relevant orders in $\epsilon$)
\begin{align}
&K_{00}|_\M=-g_H \epsilon +\O(\epsilon^3)\ ,\\
&K_{0A}|_\M=\O(\epsilon)\ , \\
&K_{AB}|_\M=\frac{1}{\epsilon}\theta_{AB}+\O(\epsilon)\ .\label{kab}
\end{align}
Therefore, the trace is given by
\begin{equation}
K\equiv h^{ij}K_{ij}=\frac{1}{\epsilon}\left(g_H+\theta_H \right)+\O(\epsilon)\ .\label{ktrace}
\end{equation}

\subsection{Viscous Newtonian fluid}\label{newton1}
The energy-momentum tensor of a 3-dimensional viscous Newtonian fluid with pressure $p$ and energy density $\rho$ is given by\footnote{Let us recall that indices $i,j$ run over all the coordinates on the stretched horizon, whereas indices $A, B$ run over only the spatial coordinates on the stretched horizon.}
\begin{equation}\label{newton}
T^{ij}=(\rho+p)u^iu^j+p h^{ij}-\tau^{ij}\ ,
\end{equation}
where, $u^i$ is the fluid velocity and $\tau^{ij}$ is the the dissipative part of the energy-momentum tensor 
\begin{equation}
\tau^{ij}=-P^{ik}P^{jl}\left[\eta f_{kl}+\zeta h_{kl}D^{(3)}_m u^m\right]\ ,
\end{equation}
where $P^{ij}=h^{ij}+u^iu^j$ is the projection operator and
\begin{equation}
f_{kl}=D^{(3)}_k u_l+D^{(3)}_l u_k-h_{kl}D^{(3)}_m u^m\ .
\end{equation}
$\eta$ is the shear viscosity and $\zeta$ is the bulk viscosity. Therefore, we can write the full energy-momentum tensor in the following way
\be\label{newton}
T^{ij}=\rho u^iu^j+ P^{ij}\left(p- \zeta D^{(3)}_m u^m\right)-\eta P^{ik}P^{jl}  f_{kl} \ .
\ee

Comparing the energy momentum tensor on the stretched horizon (\ref{membrane_em}) with (\ref{newton}) we obtain\footnote{See appendix \ref{transport} for details. We will also provide an alternative derivation of these relations in section \ref{potot}.}
\begin{align}
&\rho=-\frac{\m^2}{\epsilon} \theta_H\ , \qquad p=\frac{\m^2}{\epsilon} g_H\ ,\\
& \zeta=- \frac{\m^2}{2} \ , \qquad \eta=\frac{\m^2}{2} \label{etah}\ .
\end{align}
Stretched horizon of a black hole indeed behaves like a viscous Newtonian fluid. Few comments are in order: both energy density and pressure diverge in the limit $\epsilon\rightarrow 0$ because of the large blueshift factor near the horizon. It is more appropriate to define renormalized energy density $\rho_H= \epsilon \rho$ and renormalized pressure $p_H=\epsilon p$ which are energy and pressure as measured at infinity \cite{Price:1986yy}. One can check that for a Schwarzschild or Kerr black hole $\theta_H=0$ and hence $\rho_H=0$.  Also note that the stretched horizon has negative bulk viscosity. In ordinary fluid, negative bulk viscosity indicates instability, however, it is alright for a horizon to have negative bulk viscosity. Negative bulk viscosity of the horizon indicates that expansion of the horizon is acausal and one must impose teleological boundary conditions \cite{Price:1986yy, Thorne:1986iy}.


\section{Gravitational $\Theta$-term}\label{theta}

In any sensible theory of quantum gravity, the Einstein-Hilbert action should be the leading low energy term. However, it is expected that the low energy limit will also generate higher derivative correction terms. The only ghost-free combination,  in order $R^2$,  is the Gauss-Bonnet term.\footnote{If we allow higher derivative corrections then the Lanczos-Lovelock gravity is the unique extension of the Einstein gravity with equations of motion containing only up to two time derivatives \cite{Lovelock:1971yv}. In the Lanczos-Lovelock gravity, the Gauss-Bonnet term is the first (and in $(3+1)-$dimensions the only) correction to the Einstein gravity. The membrane paradigm  for the Lanczos-Lovelock gravity has been studied in \cite{Kolekar:2011gg}. } Effects of the Gauss-Bonnet term on the stretched horizon have already been studied in \cite{Jacobson:2011dz}. In $(3+1)-$dimensions, the Gauss-Bonnet term is a total derivative and hence does not contribute to the equations of motion. However, in the presence of this topological term the entropy of a black hole receives a correction term proportional to the Euler number of the horizon. One can show that this correction term violates the second law of black hole thermodynamics and hence should be zero \cite{Jacobson:1993xs,Olea:2005gb, Liko:2007vi, Sarkar:2010xp}. In $(3+1)-$dimensions, there can exist another total derivative term, a parity violating gravitational $\Theta$-term.\footnote{Possibility of gravitationally induced CP-violation because of the $\Theta$-term was already explored in 1980 \cite{Deser:1980kc}.} We are interested in figuring out the effects of this term on the stretched horizon which are forbidden in a parity-invariant theory.

Let us now introduce the parity violating $\Theta$-term  to the $(3+1)-$dimensional gravity action
\begin{equation}\label{CSGR}
 S=S_{EH}+S_{GHY}+S_{\Theta}=\int d^4x\sqrt{-g}\left(\frac{\m^2}{2}R +\frac{ \Theta}{4} R*R\right)+\m^2\int_\Sigma d^3 x\sqrt{|h|}K\ ,
\end{equation}
where, $\Theta$ is a constant and the quantity $R*R$ is the Chern-Pontryagin density which is defined as
\begin{equation}
R*R=R^\tau_{~\sigma \mu \nu}*R^{\sigma ~ \mu \nu}_{~\tau}
\end{equation}
where
\begin{equation}
*R^{\sigma ~ \mu \nu}_{~\tau}\equiv \frac{1}{2}e^{\mu \nu \alpha \beta}R^{\sigma}_{~\tau \alpha \beta} \ ,
\end{equation}
and $e^{\mu \nu \alpha \beta}$ is the Levi-Civita tensor.\footnote{Levi-Civita tensor $e^{\mu \nu \alpha \beta}$ is related to the Levi-Civita tensor density $\epsilon^{\mu \nu \alpha \beta}$ in the following way
\begin{equation}
e^{\mu \nu \alpha \beta}=\frac{\epsilon^{\mu \nu \alpha \beta}}{\sqrt{g}}\ .
\end{equation}
} 
Unlike the Chern-Simons modified gravity discussed in \cite{Jackiw:2003pm, Alexander:2009tp}, we are interested in the constant $\Theta$ case because for constant $\Theta$, the Chern-Pontryagin density term is a total derivative. Applying the membrane paradigm we will show that this total derivative term has non-trivial effect on the stretched horizon.

We will impose the Dirichlet boundary conditions on the boundary of space-time $\Sigma$, i.e. $\delta g^{\mu\nu}=0$ over the outer boundary of space-time. Variation of the $\Theta$-term generates boundary terms on $\Sigma$ which do not vanish with the Dirichlet boundary conditions on $\Sigma$. The action consistent with the Dirichlet boundary condition on the boundary is obtained by adding a Gibbons-Hawking-York like boundary term for the $\Theta$-term \cite{Grumiller:2008ie}
\begin{equation}\label{sbt}
S_{b\Theta}=\Theta \int_{\Sigma}d^3x\sqrt{|h|}n_\alpha e^{\alpha \beta \gamma \delta}K_\beta^{~\rho} \nabla_\gamma K_{\delta \rho}\ .
\end{equation}
The action (\ref{CSGR}) should now be replaced by the following action which is well behaved with the Dirichlet boundary condition
\begin{equation}\label{stot}
 S=\int d^4x\sqrt{-g}\left(\frac{\m^2}{2}g^{\mu \nu}R_{\mu \nu} +\frac{\Theta}{8}e^{\mu \nu \alpha \beta}R^\tau_{~\sigma \mu \nu} R^{\sigma}_{~\tau \alpha \beta}\right)+S_{GHY}(\Sigma)+S_{b\Theta}(\Sigma)\ .
\end{equation}
The $\Theta$-term does not affect the classical equations of motion because it is a total derivative. However, similar to the electrodynamics $\theta$-term \cite{Fischler:2015cma},  the gravitational $\Theta$-term also generates non-trivial boundary terms which can affect the dynamics of the stretched horizon.

\subsection{Energy-momentum tensor on the stretched horizon}
Let us again imagine a stretched horizon $\M$ that divides the whole space-time in two regions
\begin{align}
A:&~\qquad \text{outside the membrane}~ \M \ ,\nonumber\\
B:&~\qquad \text{inside the membrane}~ \M\ \nonumber 
\end{align}
and an observer $\O$ who has access only to the region outside the membrane. Following the discussion of the previous section we can write down the path integral in the following way
 \be\label{bcs1}
Z=\int_{A+\M} \D[g_{\mu\nu}^\M]\D[g_{\mu\nu}^A] e^{i \left(S_A-S_{GHY}(\M)-S_{b\Theta}(\M)\right)}\int_B\D[g_{\mu\nu}^B]e^{i \left(S_B+S_{GHY}(\M)+S_{b\Theta}(\M)\right)}\ ,
\ee
where, the total action (\ref{stot}) has been divided as $S=S_A+S_B$. Now we fix the metric on $\M$ and then write  the path integral on $B$ as
\be\label{so}
\int_B\D[g_{\mu\nu}^B]e^{i \left(S_B+S_{GHY}(\M)+S_{b\Theta}(\M)\right)}=e^{i \S_{0}(\M)}\ .
\ee
Hence, the partition function (\ref{bcs1}) becomes,
 \be\label{bcs2}
Z_\O=\int_{A+\M} \D[g_{\mu\nu}]e^{i \left(S_A-S_{b\Theta}(\M)-S_{GHY}(\M)+\S_{0}(\M)\right)}\ ,
\ee
where, now the path integral is defined only on or outside of the stretched horizon $\M$. The physics outside the black hole classically must not be affected by the dynamics inside the black hole. That means the observer $\O$ should be able to obtain the correct classical equations of motion by varying action $S_\O$ which is restricted only to the space-time outside the black hole, where,\footnote{Note that we are using the notation $\int^{\Sigma}_{\M} d^3 x\sqrt{|h|}K=\int_{\Sigma} d^3 x\sqrt{|h|}K-\int_{\M} d^3 x\sqrt{|h|}K$.}
\begin{align}\label{csgravity}
 S_\O=&\int_A d^4x\sqrt{-g}\left(\frac{\m^2}{2}g^{\mu \nu}R_{\mu \nu} +\frac{\Theta}{8}e^{\mu \nu \alpha \beta}R^\tau_{~\sigma \mu \nu} R^{\sigma}_{~\tau \alpha \beta}\right)+\m^2\int^{\Sigma}_{\M} d^3 x\sqrt{|h|}K\nonumber\\
&~~~~~~~~~~~~~~~~~~~~~~~~+\Theta \int^{\Sigma}_\M d^3x\sqrt{|h|}n_\alpha e^{\alpha \beta \gamma \delta}K_\beta^{~\rho} \nabla_\gamma K_{\delta \rho}+\S_{0}(\M)\nonumber \\
=&S_{EH}+S_{\Theta}+S_{GHY}(\Sigma)-S_{GHY}(\M)+S_{b\Theta}(\Sigma)-S_{b\Theta}(\M)+\S_{0}(\M)\ .
\end{align}
Before we proceed note that $S_{\Theta}+S_{b\Theta}$ is equivalent to the three-dimensional gravitational Chern-Simons term\cite{Grumiller:2008ie}
\begin{align}
&\int d^4x\sqrt{-g} \frac{\Theta}{4} R*R +\Theta \int_{\Sigma}d^3x\sqrt{|h|}n_\alpha e^{\alpha \beta \gamma \delta}K_\beta^{~\rho} \nabla_\gamma K_{\delta \rho}~~~~~\nonumber\\
&~~~~~~~~~~~~~~~~~~~~~~~~=\Theta \int_{\Sigma}d^3x \sqrt{|h|} e^{ijk} \left[\frac{1}{2}\gamma^l_{im}\partial_j \gamma^m_{kl}+\frac{1}{3}\gamma^l_{im}\gamma^m_{jp}\gamma^p_{kl}\right]\label{3cs}
\end{align}
where
\begin{equation}
e^{ijk}\equiv n_\mu e^{\mu ijk}\ .
\end{equation}
The three-dimensional action (\ref{3cs}) is exactly the gravitational Chern-Simons action in $(2+1)$-dimensions. Now the action for the observer $\O$ is
\begin{align}
 S_\O=&\frac{\m^2}{2}\int_A d^4x\sqrt{-g}g^{\mu \nu}R_{\mu \nu} +\m^2\int^{\Sigma}_{\M} d^3 x\sqrt{|h|}K\nonumber\\
&~~~~~~~~~~~~~~+\Theta \int^{\Sigma}_\M d^3x \sqrt{|h|} e^{ijk} \left[\frac{1}{2}\gamma^l_{im}\partial_j \gamma^m_{kl}+\frac{1}{3}\gamma^l_{im}\gamma^m_{jp}\gamma^p_{kl}\right]+\S_{0}(\M)\ .
\end{align}
Variation of the gravitational Chern-Simons action is well known 
\begin{equation}\label{varcs}
\delta  \int^{\Sigma}_\M d^3x \sqrt{|h|} e^{ijk} \left[\frac{1}{2}\gamma^l_{im}\partial_j \gamma^m_{kl}+\frac{1}{3}\gamma^l_{im}\gamma^m_{jp}\gamma^p_{kl}\right]=-\int^{\Sigma}_\M d^3x \sqrt{|h|} C_{mn}\delta h^{mn}
\end{equation}
where $C_{mn}$ is the Cotton-York tensor
\begin{equation}\label{cy}
C_{mn}=-\frac{e^{ijk}}{2}\left[h_{mk} D^{(3)}_{i} R^{(3)}_{nj}+h_{nk} D^{(3)}_{i} R^{(3)}_{mj} \right]\ ,
\end{equation}
where, $R^{(3)}_{ij}$ is the Ricci tensor of the three-dimensional surface $\M$ (or $\Sigma$) and $D^{(3)}_{i}$ is the three-dimensional covariant derivative. The Cotton-York tensor $C_{mn}$ is symmetric, traceless and covariantly conserved in three-dimensions. Therefore, the three-dimensional stretched horizon energy-momentum tensor receives a correction because of the gravitational $\Theta$-term
\begin{equation}\label{theta_em}
 \T_{i j}=\m^2\left[h_{i j}K -K_{ij}\right]|_\M - 2 \Theta C_{ij}|_\M\ .
\end{equation}

\subsection{Parity odd third order transport}\label{potot}

We already know that a black hole horizon behaves like a fluid for an observer hovering outside. In section (\ref{newton1}), we have derived different first order transport coefficients of this fluid for Einstein gravity.  Let us now figure out how the $\Theta$-term in (\ref{theta_em}) affects the transport property of the fluid living on a stretched horizon. Transport coefficients are the measure of the response of a fluid after hydrodynamic perturbations. For example, one can disturb a hydrodynamic system by perturbing the background metric and then observe the change in the energy momentum tensor of the fluid as a result of this perturbation. In order to do that we will study linearized perturbation of the metric near the horizon of a stationary black hole solution. Again we consider an observer $\O$ hovering just outside a $(3+1)-$dimensional black hole. For such an observer, the unperturbed near horizon metric is Rindler-like
\be\label{up}
ds^2=-r^2~d\t^2+ \frac{2r}{g_H}d\t dr+ dx_1^2+dx_2^2\ .
\ee
For the observer $\O$, there is a horizon at the edge of the Rindler wedge $r=0$. Note that we are restricting to the case  of constant $g_H$ on the horizon. Let us now perturb the metric: $g_{\mu \nu}=g^0_{\mu \nu}+g^1_{\mu\nu}$, where $g^0_{\mu \nu}$ is the unperturbed metric (\ref{up}). We perturb the metric such that very close to the horizon only $g^1_{AB}\equiv \gamma_{AB}(\t,r,x^1,x^2)$ components are non-zero
\be\label{pm}
ds^2=-r^2~d\t^2+ \frac{2r}{g_H}d\t dr+ dx_1^2+dx_2^2+ \gamma_{AB}(\t,r,x^1,x^2) dx^A dx^B\ ,
\ee
where indices $A, B=1,2$ run over the transverse spatial directions. Let us now use (\ref{theta_em}) to find out how the energy-momentum tensor of the horizon fluid responds to this perturbation.

From equations (\ref{kab}) and (\ref{ktrace}), we find that the extrinsic curvature at the stretched horizon (at $r=\epsilon$) in linear order in perturbations, is given by
\be\label{k1}
K_{AB}|_\M=\frac{1}{2\epsilon}\left(\frac{\partial \gamma_{AB}}{\partial \t}\right)\ , \qquad K|_\M=\frac{1}{\epsilon}\left(g_H+\frac{1}{2}\frac{\partial \gamma_{11}}{\partial \t}+\frac{1}{2}\frac{\partial \gamma_{22}}{\partial \t}\right)\ ,
\ee
where, we have ignored terms with positive powers of $\epsilon$. Similarly, we can calculate the Cotton-York tensor (\ref{cy}) on the stretched horizon. In linear order in perturbations and in the leading order in $\epsilon$, we obtain  
\begin{align}
&C_{11}=-C_{22}=\frac{1}{2 \epsilon^3} \frac{\partial^3 \gamma_{12}}{\partial \t^3}\ ,\label{C1}\\
&C_{12}=C_{21}=-\frac{1}{4 \epsilon^3}\left( \frac{\partial^3 \gamma_{11}}{\partial \t^3}- \frac{\partial^3 \gamma_{22}}{\partial \t^3}\right) \label{C2}
\end{align}
and all the other components vanish. Before we proceed, we want to note that the Cotton-York tensor contains derivatives only with respect to the coordinates on the stretched horizon. Hence, we do not need to solve the Einstein equations in order to write down the $\Theta$-contributions to the energy-momentum tensor in terms of quantities intrinsic to the stretched horizon.  

Let us now study the effect of the metric perturbation (\ref{pm}), order by order,  on the energy-momentum tensor on the stretched horizon. In the $0^{\text{th}}$ order in derivative expansion, from (\ref{theta_em}), we obtain
\be
\T_{00}=\O(\epsilon^3)\ , \qquad \T_{AB}= \frac{\m^2 g_H}{\epsilon} \left(\delta_{AB}+\gamma_{AB}\right)\ .
\ee 
Comparing the $0^{\text{th}}$ order energy-momentum tensor with that of an ideal fluid, we find that energy density $\rho=0$ and pressure $p=\m ^2 g_H/\epsilon$. The renormalized pressure which is the pressure as measured at infinity, is given by $p_H = \m ^2 g_H$.

In the $1^{\text{st}}$ order in derivative expansion, from (\ref{theta_em}) and (\ref{k1}), we obtain
\begin{align}
&\T_{11}=\frac{\m^2}{2\epsilon}\left( \frac{\partial \gamma_{22}}{\partial \t}\right)\ , \qquad \T_{22}=\frac{\m^2}{2\epsilon}\left( \frac{\partial \gamma_{11}}{\partial \t}\right)\ ,\label{first1}\\
&\T_{12}=\T_{21}=-\frac{\m^2}{2\epsilon}\left( \frac{\partial \gamma_{12}}{\partial \t}\right)\ .\label{first2}
\end{align}
For a Newtonian viscous fluid in $(2+1)-$dimensions, in the $1^{\text{st}}$ order in derivative expansion, the energy momentum tensor has the form:
\begin{align}
T_{AB}=-\eta \frac{\partial \gamma_{AB}}{\partial t}-\zeta \delta_{AB}\left(\frac{\partial \gamma_{11}}{\partial t}+\frac{\partial \gamma_{22}}{\partial t}\right)\ ,
\end{align}
where we have assumed that the spatial part of the unperturbed metric is flat. Hence, the stretched horizon, for an observer hovering outside, has shear viscosity $\eta=\m^2/2$ and bulk viscosity $\zeta=-\m^2/2$, which agrees with (\ref{etah}). Note that the $1^{\text{st}}$ order energy-momentum tensor (\ref{first1}-\ref{first2}) diverges in the limit $\epsilon\rightarrow 0$ because of large blueshift factor near the horizon. One should think of equations (\ref{first1}-\ref{first2}) as variations of the metric on the stretched horizon with respect to time at infinity: $t=\epsilon\ \t$.

Energy-momentum tensor on the stretched horizon (\ref{theta_em}) does not contain any term in the $2^{\text{nd}}$ order in derivative expansion. However, from equations (\ref{C1}-\ref{C2}), it is clear that the $\Theta$-term contributes in the $3^{\text{rd}}$ order in derivative expansion, yielding 
\begin{align}
&\T_{11}=-\T_{22}=-\frac{\Theta}{ \epsilon^3} \frac{\partial^3 \gamma_{12}}{\partial \t^3}\ ,\label{TC1}\\
&\T_{12}=\T_{21}=\frac{\Theta}{2 \epsilon^3}\left( \frac{\partial^3 \gamma_{11}}{\partial \t^3}- \frac{\partial^3 \gamma_{22}}{\partial \t^3}\right) \ . \label{TC2}
\end{align}
Let us now define a new third order, parity violating transport coefficient  in $(2+1)-$dimensions $\vartheta$ which contributes to the energy-momentum tensor in the following way\footnote{We will discuss about the transport coefficient $\vartheta$ in more details in section \ref{section_kubo}.}
\begin{align}
&T_{11}=-T_{22}=-\vartheta \frac{\partial^3 \gamma_{12}}{\partial t^3}\ ,\label{third1}\\
&T_{12}=T_{21}=\frac{\vartheta}{2 }\left( \frac{\partial^3 \gamma_{11}}{\partial t^3}- \frac{\partial^3 \gamma_{22}}{\partial t^3}\right) \ . \label{third2}
\end{align}
Hence, the stretched horizon, for an observer hovering outside, has
\be
\vartheta=\Theta\ .
\ee
Note that this is not Hall viscosity but a {\it third order cousin} of Hall viscosity.\footnote{Hall viscosity contributes in the first order in derivative expansion:
\begin{align}
&T_{11}=-T_{22}=-\eta_A \frac{\partial \gamma_{12}}{\partial t}\ ,\\
&T_{12}=T_{21}=\frac{\eta_A}{2 }\left( \frac{\partial \gamma_{11}}{\partial t}- \frac{\partial \gamma_{22}}{\partial t}\right) \ . 
\end{align}
} 
The linear response theory dictates that the retarded Green's functions of the energy-momentum tensor on the stretched horizon are
\begin{align}
G^R_{12,11}(\omega,k)=-G^R_{12,22}(\omega, k)=-\frac{i \Theta \omega^3}{ \epsilon^3}\ ,
\end{align}
where, the presence of $1/\epsilon^3$ factor is again a consequence of large blueshift near the horizon. Therefore, the gravitational $\Theta$-term affects the transport property of the stretched horizon and in principle it can be found from the coefficient of the $\omega^3$ term of  $G^R_{12,11-22}(\omega,k)$. 

\subsection{Fast scrambling}
We will end this section with a discussion of the effect of the gravitational $\Theta$-term on the way perturbations scramble on the horizon. The process by which a localized perturbation spreads out into the whole system is known as {\it scrambling}. In quantum mechanics, information contained  inside a small subsystem of a larger system  is fully scrambled when the small subsystem becomes entangled with the rest of the system and after scrambling time $t_s$  the information can only be retrieved by examining practically all the degrees of freedom. In a local quantum field theory, it is expected that the scrambling time $t_s$ is at least as long as the diffusion time. Consequently, one can show that the scrambling time for a strongly correlated quantum fluid in $d$-spatial dimensions and at temperature $T$, satisfies 
\be\label{bound}
t_s T \ge c \hbar S^{2/d}\ ,
\ee
where $c$ is some dimensionless constant and $S$ is the total entropy. In \cite{Sekino:2008he,Susskind:2011ap}, it has been argued that (\ref{bound}) is a universal bound on the scrambling time. Hence, it is indeed remarkable, as shown in  \cite{Sekino:2008he,Susskind:2011ap}, that information scrambles on black hole horizons exponentially fast
\be
t_s T \approx \hbar \ln S
\ee
violating the bound (\ref{bound}). This unusual process is known as ``fast-scrambling" and it strongly suggests that the microscopic description of fast-scrambling on black hole horizons must involve non-local degrees of freedom \cite{Sekino:2008he,Susskind:2011ap}. It is now well known that non-locality is indeed essential for fast scrambling \cite{nonlocal1,nonlocal2,nonlocal3,nonlocal4,nonlocal5} which is also supported by the observation that non-local interactions can enhance the level of entanglement among different degrees of freedom \cite{entanglement1,entanglement2,entanglement3}.

Recently it has been shown that in the presence of the electrodynamics $\theta$-angle a black hole horizon behaves as a Hall conductor, for an observer hovering outside \cite{Fischler:2015cma}. As a consequence, the electrodynamics $\theta$-angle affects the way localized perturbations on the stretched horizon, created by dropping a charged particle, Hall-scramble on the horizon \cite{Fischler:2015cma}. 

In presence of the gravitational $\Theta$-term, it is reasonable to expect a similar conclusion. We can perform a thought experiment, in which an outside observer drops a massive particle onto the black hole and watches how the perturbation scrambles on the black hole horizon. Equations (\ref{third1}-\ref{third2}) indicate that the gravitational $\Theta$-term will also affect the way perturbations scramble on the horizon, in particular, it will introduce vortices without changing the scrambling time. This perhaps suggests that a microscopic description of fast-scrambling needs to be able to explain the origin of this effect.\footnote{It will be very interesting to explore if microscopic models of fast-scrambling, such as \cite{Barbon:2012zv, Barbon:2013goa} can be used to implement these parity violating scrambling processes.} Following \cite{Fischler:2015cma} it is possible to perform an explicit calculation for this effect but we will not attempt it in this paper. 

It is important to note that the Rindler approximation of the near horizon metric (\ref{up}) has a crucial limitation. In a Schwarzschild black hole, any freely falling object will hit the singularity in finite  proper time, the effect of which on the stretched horizon can not be analyzed in the near horizon approximation. However, one can argue  that  when a massive particle hits the singularity, the spherical symmetry will be restored \cite{Banks:2014xja}. This should not be surprising because an order one perturbation will decay to size $ \sim M_P/m$ in one scrambling time $t_s$, where $m$ is the mass of the black hole \cite{Sekino:2008he}. Hence, this should be the time scale for any classical fields on a spherically symmetric horizon to become spherically symmetric.

\section{Gauge/gravity duality}\label{adscft}
The AdS/CFT correspondence \cite{Maldacena:1997re, Witten:1998qj, Gubser:1998bc, Aharony:1999ti} has successfully provided us with theoretical control over a large class of strongly interacting field theories. It is indeed remarkable that observables of certain large-N gauge theories in $d$-dimensions can be calculated by performing some classical gravity computations in $(d+1)$-dimensions. Gravity duals of these field theories at finite temperature contain black holes, where the field theory temperature is given by the Hawking temperature of black holes. It is well known that at very long length scales the most dominant contributions to different non-local observables come from the near horizon region of the dual black hole geometry \cite{Fischler:2012ca, Fischler:2012uv}. So, it is not very surprising that there is some connection between the low energy hydrodynamic description of a strongly coupled quantum field theory and the membrane paradigm fluid on the black hole horizon of the dual gravity theory \cite{Kovtun:2003wp,Kovtun:2004de,Son:2007vk,Iqbal:2008by,Bredberg:2010ky}. Long before the emergence of gauge/gravity duality, it was shown that the membrane paradigm fluid on the stretched horizon of a black hole has a shear viscosity to entropy density ratio of $1/4\pi$ \cite{Price:1986yy, Thorne:1986iy}. Later it was found that the shear viscosity to entropy density ratio of a gauge theory with a gravity dual is indeed $1/4\pi$ \cite{Kovtun:2004de,Bredberg:2010ky}. Interestingly, the strong coupling physics of quark-gluon plasma has been experimentally explored in the Relativistic Heavy Ion Collider (RHIC), where the shear viscosity to entropy density has been measured to be close to $1/4\pi$. On the other hand, it is also known that there are membrane paradigm results which differ from the AdS/CFT values. For example, the bulk viscosity of a conformal fluid is exactly zero, whereas  the membrane paradigm bulk viscosity is not only nonzero but negative.\footnote{See \cite{deBoer:2014xja} for another example when membrane paradigm is incomplete in the context of the AdS/CFT correspondence.} 

\subsection{Energy-momentum tensor}
Let us now consider a large-N gauge theory in $(2+1)-$dimensions which is dual to a gravity theory in $(3+1)-$dimensions with the gravitational $\Theta$-term
\begin{equation}\label{CSGR}
 S=\int d^4x\sqrt{-g}\left[\frac{\m^2}{2}\left(R -\frac{6}{L^2}\right) +\frac{ \Theta}{4} R*R\right]\ ,
\end{equation}
where, cosmological constant is $-3/L^2$. Note that this is different from the case considered in \cite{Saremi:2011ab}, where $\Theta$ was dynamical.\footnote{In \cite{Saremi:2011ab}, it was also assumed that $\Theta$ vanishes asymptotically. } In the presence of a dynamical $\Theta(r)$, the boundary theory exhibits Hall viscosity,\footnote{In (2+1) dimensions, the Hall viscosity contributes to the energy-momentum tensor in the first order in derivative expansion
\begin{equation}
T_H^{ij}=-\frac{\eta_{A}}{2}\left(e^{ikl}u_k f_l^j+e^{jkl}u_k f_l^i\right)\ ,
\end{equation}
where, $u^i$ is the fluid velocity,  $e^{ikl}$ is the Levi-Civita tensor and $f_{kl}=D^{(3)}_k u_l+D^{(3)}_l u_k-h_{kl}D^{(3)}_m u^m$.}
where $\eta_A$ is proportional to $\Theta'(r)$ at the horizon $r=r_H$ \cite{Saremi:2011ab}.\footnote{Hall viscosity has been studied extensively in the context of Holography, e.g. see \cite{Liu:2014gto,Chen:2011fs,Chen:2012ti,Cai:2012mg,Zou:2013fua,Wu:2013vya,Roychowdhury:2014jqa,Son:2013xra}} However, we are interested in the constant $\Theta$ case for which the $\Theta$-term is a total derivative and hence $\eta_A=0$. 

Now we figure out the effect of the parity violating $\Theta$-term on the dual field theory. A reasonable guess is that the boundary theory, similar to the membrane paradigm fluid, will acquire the third order parity violating ``Hall viscosity-like" transport coefficient $\vartheta$ as defined in equations (\ref{third1}-\ref{third2}), with $\vartheta=\Theta$.  We will confirm this guess by performing an explicit computation. We will also show that in the presence of the gravitational $\Theta$-term, the two-point function of the energy-momentum tensor of the boundary theory acquires a non-trivial contact term. The fractional part of this contact term does not depend on the short distance physics and hence it is a meaningful physical observable in $(2+1)-$dimensional conformal field theory.

The action (\ref{CSGR}) is not consistent with Dirichlet boundary condition, so we are required to add boundary terms for the variational principle  to be well defined
\begin{equation}
 S=\int d^4x\sqrt{-g}\left[\frac{\m^2}{2}\left(R -\frac{6}{L^2}\right) +\frac{ \Theta}{4} R*R\right]+S_{GHY}+S_{b\Theta}+S_{ct}\ .
\end{equation}
$S_{b\Theta}$ is a Gibbons-Hawking-York like boundary term for the $\Theta$-term, defined in equation (\ref{sbt}). Counterterm $S_{ct}$ that we need to add in order to get a finite boundary energy momentum tensor  \cite{Henningson:1998gx, Balasubramanian:1999re} is the constructed with quantities intrinsic to the boundary geometry. For $(3+1)-$dimensions, $S_{ct}$ is given by \cite{Balasubramanian:1999re}
\be
S_{ct}=\frac{2\m^2}{L}\int d^3x \sqrt{-h}\left(1-\frac{L^2}{4} R^{(3)}\right)\ .
\ee

We are mainly interested in the contribution of the $\Theta$-term to the boundary energy momentum tensor. From the membrane paradigm calculation (\ref{varcs}), we already know that 
\begin{equation}
\delta  \left(\int d^4x\sqrt{-g}\frac{ \Theta}{4} R*R+S_{b\Theta}\right)=-\int_{\Sigma} d^3x \sqrt{|h|} C_{mn}\delta h^{mn}
\end{equation}
where $C_{mn}$ is the Cotton-York tensor (\ref{cy}). Therefore, following \cite{Balasubramanian:1999re}, the $\Theta$-term contribution to the energy-momentum tensor of the dual field theory is given by\footnote{Similar observation was also made by \cite{Miskovic:2009bm}.}
\begin{equation}\label{membrane_em1}
 T^\Theta_{i j}=\lim_{r_b\rightarrow \infty}\left(\frac{-2 r_b \Theta C_{ij}}{L}\right)_{boundary}\ ,
\end{equation}
where, the boundary of the asymptotically AdS$_{(3+1)}$ is at $r=r_b$.\footnote{Note that in \cite{Mukhopadhyay:2013gja}, authors constructed $(3+1)-$dimensional bulk geometries for which the boundary Cotton-York tensor has the form of the energy momentum tensor of a perfect fluid. These solutions have non-trivial boundary geometries which lead to interesting effects even in equilibrium. However, we are interested in asymptotically AdS$_{(3+1)}$ spacetime where the $\Theta$-term only affects the out of equilibrium dynamics.}

\subsection{Linearized perturbations}
Let us now consider the asymptotically AdS$_{(3+1)}$ spacetime,
\be\label{ads}
ds^2=2H(r)dvdr-\frac{r^2}{L^2} f(r)dv^2+\frac{r^2}{L^2}\left(dx_1^2+dx_2^2 \right)\ .
\ee
where, both $f(r)$ and $H(r)$ go to 1 near the boundary. One can easily check that for this metric,  $R*R$ vanishes. Now we will perturb the metric in the following way,
\be
ds^2=2H(r)dvdr-\frac{r^2}{L^2} f(r)dv^2+\frac{r^2}{L^2}\left(dx_1^2+dx_2^2 \right)+\frac{r^2}{L^2}h_{AB}dx^Adx^B\ ,
\ee
where $A,B=1,2$ and
\be
h_{12}(r,v)=h_{21}(r,v)\ , \qquad h_{11}(r,v)=-h_{22}(r,v)\ .
\ee
One can easily check that this set of perturbations decouple from the rest of the components. Now using equation (\ref{membrane_em1}), we obtain the contribution of the $\Theta$-term on the energy-momentum tensor of the $(2+1)-$dimensional boundary theory
\begin{align}
&T^\Theta_{11}=- \frac{\Theta}{f(r_b)^{3/2}}\left(\frac{\partial^3 h_{12}}{\partial v^3}\right)_{boundary}\ ,\\
&T^\Theta_{12}=T^\Theta_{21}= \frac{\Theta}{2f(r_b)^{3/2}}\left(\frac{\partial^3 h_{11}}{\partial v^3}-\frac{\partial^3 h_{22}}{\partial v^3}\right)_{boundary}\ ,\\
&T^\Theta_{22}= \frac{\Theta}{f(r_b)^{3/2}}\left(\frac{\partial^3 h_{12}}{\partial v^3}\right)_{boundary}\ .
\end{align}
Note that we have obtained the energy-momentum tensor off-shell because we do not need to solve the Einstein equations in order to write down the $\Theta$-contributions to the energy-momentum tensor.

Fluctuations of the boundary metric is given by: $\gamma_{AB}=h_{AB}(r=r_b)$. In the boundary theory, we are considering zero-momentum modes: $\gamma_{AB}\sim e^{-i\omega t}$, where at the boundary $v=t$ and hence
\be\label{eqn1234}
T^\Theta_{11}=-T^\Theta_{22}=-\Theta \frac{\partial^3 \gamma_{12}}{\partial t^3}\ , \qquad T^\Theta_{12}=T^\Theta_{21}=\frac{\Theta}{2} \left(\frac{\partial^3 \gamma_{11}}{\partial t^3}-\frac{\partial^3 \gamma_{22}}{\partial t^3}\right)\ .
\ee
In the last equation, we have used the fact that both $f(r)$ and $H(r)$ go to 1 near the boundary. Therefore, comparing equation (\ref{eqn1234}) with (\ref{third1}-\ref{third2}), we find that the $(2+1)$-dimensional boundary theory has a nonzero $\vartheta$, in particular  
\be
\vartheta= \Theta\ ,
\ee
which agrees with the membrane paradigm result. Interestingly, for a holographic theory $\vartheta$ is independent of the temperature. This is a consequence of the fact that the $\Theta$-contribution to the boundary theory energy-momentum tensor is always a local  functional of the boundary metric and does not depend on the interior geometry. This also strongly indicates that the transport coefficient $\vartheta$ is completely independent of the quantum state of the field theory. It is important to note that the most transport coefficients make sense only at finite temperature and in the low frequency limit $\omega/T\rightarrow 0$. However, this is not true for this new transport coefficient $\vartheta$, which as we will show next arises from a state-independent contact term in the energy-momentum tensor and hence in some sense is only probing the physics of the vacuum.

\subsection{Contact term in the two-point function}
We now compute the two-point function of the energy-momentum tensor of a $(2+1)-$dimensional quantum field theory dual to a gravity theory in $(3+1)-$dimensions with the gravitational $\Theta$-term (\ref{CSGR}).\footnote{We are grateful to T. Hartman for a discussion that led to this subsection.} We again consider asymptotically AdS$_{(3+1)}$ spacetime (\ref{ads}) and calculate the contribution of the $\Theta$-term to the two-point function of the energy-momentum tensor. We perturb the metric: $g_{\mu\nu}+\delta g_{\mu \nu}$, such that $\delta g_{rr}=\delta g_{r i}=0$ at the boundary. The on-shell action $(S_\Theta+S_{b\Theta})$ in the second order in metric perturbations is obtained to be\footnote{In this subsection, we will be working in the Euclidean signature and we will use the convention $\gamma_{ij}(x)=\frac{1}{(2\pi)^{3/2}}\int d^3k\ e^{ip.x}\gamma_{ij}(p)$ for the Fourier transform.}
\be
S_\Theta+S_{b\Theta}=\frac{\Theta}{16} \int d^3p\ \T_{ij;mn}\gamma^{ij}(p)\gamma^{mn}(-p)
\ee
where, $\gamma_{ij}=\frac{L^2}{r^2}\delta g_{ij}$ is the boundary metric perturbation and
\be\label{tijmn}
\T_{ij;mn}=\left(\varepsilon_{iml}p^l\left(p_j p_n-p^2 \delta_{jn} \right)+(i\leftrightarrow j)\right)+(m\leftrightarrow n)\ .
\ee
Following the AdS/CFT dictionary, we can obtain the two-point function of the boundary energy-momentum tensor by varying the above quadratic action with respect to $\gamma^{ij}$:
\be\label{corfn}
\langle T_{ij}(p)T_{mn}(-p)\rangle_\Theta=\frac{\Theta}{2}\ \T_{ij;mn}\ .
\ee
One can easily check that this gives rise to a conformally invariant contact term \cite{Closset:2012vp}
\be\label{contact}
\langle T_{ij}(x)T_{mn}(0)\rangle_\Theta=-i\frac{\Theta}{2} \left[\left(\varepsilon_{iml}\partial^l\left(\partial_j \partial_n-\partial^2 \delta_{jn} \right)+(i\leftrightarrow j)\right)+(m\leftrightarrow n)\right]\delta^3(x)\ .
\ee

It was shown in \cite{Closset:2012vp} that the two-point function of the energy-momentum tensor in a $(2+1)-$dimensional conformal field theory must have the following form
\begin{align}\label{corfn1}
\langle T_{ij}(p)T_{mn}(-p)\rangle=-\frac{\tau_g}{|p|}\left(p_{ij}p_{mn}+p_{im} p_{jn} +p_{jm}p_{in}\right)+\frac{\kappa_g}{192\pi}\T_{ij;mn}\ ,
\end{align}
where, $p_{ij}=(p_i p_j-p^2 \delta_{ij})$ and $\T_{ij;mn}$ is given in (\ref{tijmn}). The term proportional to $\tau_g$ controls the behavior of the correlation function at separated points, whereas, the term proportional to $\kappa_g$ leads to a pure contact term of the form (\ref{contact}). The coefficient $\kappa_g$ can take up any value, however, it is possible to shift $\kappa_g\rightarrow \kappa_g+ \delta \kappa_g$ by adding a gravitational Chern-Simons counterterm with coefficient $\delta \kappa_g$ to the UV Lagrangian. The gravitational Chern-Simons term, as argued in \cite{Closset:2012vp}, is a valid counterterm only if $\delta \kappa_g$ is an integer. Therefore, the integer part of $\kappa_g$ is scheme-dependent, however, the fractional part $\kappa_g$ mod $1$  does not depend on the short distance physics and hence it is a meaningful physical observable in $(2+1)-$dimensional conformal field theory \cite{Closset:2012vp}. By comparing, (\ref{corfn1}) with (\ref{corfn}), we conclude that a gravity theory in AdS$_{(3+1)}$  with the gravitational $\Theta$-term is dual to a conformal field theory with  non-vanishing $\kappa_g$, in particular
\be
\frac{\kappa_g}{96\pi}=\Theta=\vartheta\ .
\ee
This also suggests that $\Theta$ can take up any value, however, only a fractional part of the $\Theta$-parameter is a well-defined observable. It is important to note that our result is a nonlinear generalization of the contact term (\ref{contact}) and it is straight forward to obtain the contribution of the $\Theta$-term to the higher-point functions of the energy-momentum tensor of the boundary field theory.

\section{Some comments on the transport coefficient $\vartheta$: Kubo formula}\label{section_kubo}

Before we conclude, let us make some comments on the transport coefficient $\vartheta$. It is a new third order, parity violating transport coefficient  in $(2+1)-$dimensions which under a small metric perturbation $\gamma_{AB}$ around flat Minkwoski metric, contributes to the energy-momentum tensor (in the low momentum limit) in the following way:
\begin{align}
&T_{11}=-T_{22}=-\vartheta \frac{\partial^3 \gamma_{12}}{\partial t^3}\ ,\label{theta1}\\
&T_{12}=T_{21}=\frac{\vartheta}{2 }\left( \frac{\partial^3 \gamma_{11}}{\partial t^3}- \frac{\partial^3 \gamma_{22}}{\partial t^3}\right) \label{theta2}\ . 
\end{align}
Note that $\vartheta$ is dimensionless and it does not contribute to the trace of the energy-momentum tensor. It is also related to the contact term $\kappa_g$, in particular $\vartheta=\kappa_g/96\pi$. Very little is known about third-order transport coefficients in any dimensions and the transport coefficient $\vartheta$ to our knowledge has never been studied before. This is a nice example where gravity teaches us about a new hydrodynamic effect. 

In $(2+1)-$dimensional hydrodynamics, parity violating effect can appear in the first order in derivative expansion \cite{Jensen:2011xb}. Hall viscosity is an example of such effect and it has been studied for both relativistic \cite{Saremi:2011ab, Son:2013xra} and non-relativistic systems \cite{Avron:1995fg, Avron1997, Read:2010epa}. We believe that $\vartheta$ is a {\it third order cousin} of Hall viscosity and hence it should also contribute to Berry-like transport \cite{Haehl:2014zda}.

Let us now derive the Kubo formula for $\vartheta$. First, note that \cite{Son:2013xra}
\be
\langle T_{ij}(x)\rangle_\gamma=\langle T_{ij}(x)\rangle_{\gamma=0}-\frac{1}{2}\int d^3x' G^{R}_{ij,kl}(x,x')\gamma^{kl}(x')+\O(\gamma^2)\ ,
\ee
where, $i,j,k,l=0,1,2$ run over all the coordinates and $G^{R}_{ij,kl}(x,x')$ is the retarded Green's function of energy-momentum tensor
\be
G^{R}_{ij,kl}(x,x')=-i \theta(t-t')\langle\left[T_{ij}(x), T_{kl}(x')\right]\rangle\ .
\ee
Similarly, one can define the retarded Green's function in the momentum space simply by performing a Fourier transformation
\be
G^{R}_{ij,kl}(\omega, \vec{k})=\int d^3 x e^{i\omega t-i\vec{k}.\vec{x}}G^{R}_{ij,kl}(x,0)\ .
\ee
Therefore, from (\ref{theta1}-\ref{theta2}), we find that $\vartheta$ contributes to the retarded Green's function in order $\omega^3$:
\begin{align}\label{kubo}
G^R_{12,11-22}(\omega,\vec{k}\rightarrow 0)=-2i \vartheta \omega^3\ 
\end{align}
which gives the Kubo formula for $\vartheta$. Note that the Kubo formula (\ref{kubo}) can also be derived directly from equation (\ref{corfn1}).

We end this section by commenting on the possible covariant structure of the $\vartheta$-contribution to the energy-momentum tensor. It is reasonable to guess that the $\vartheta$-contribution to the energy-momentum tensor in the Landau gauge has the following form:
\be\label{tv}
T^{ij}_\vartheta=-2\vartheta \left(P^{ik}P^{j l} C_{kl}-\frac{1}{2}P^{ij} P^{kl}C_{kl}\right)\ 
\ee
with $P^{ij}=g^{ij}+u^i u^j$, where $g_{ij}$ is the $(2+1)-$dimensional metric. One can easily check that the energy-momentum tensor (\ref{tv}) under a small metric perturbation $\gamma_{AB}$ around flat Minkwoski metric leads to (\ref{theta1}-\ref{theta2}) and hence will reproduce the retarded Green's function formula (\ref{kubo}). It will be nice to derive (\ref{tv}) directly by using the formalism of fluid/gravity correspondence, however we will not attempt it here.

\section{Conclusions}\label{conclude}

We have shown that the gravitational $\Theta$-term can have physical effect on the horizon of a black hole in $(3+1)$-dimensions.  In particular,  in the presence of the $\Theta$-term, the horizon acquires a third order parity violating, dimensionless transport coefficient $\vartheta$, which affects the way localized perturbations scramble on the horizon. This strongly suggests that a sensible theory of quantum gravity should be able to provide a microscopic description of this effect. It will be very interesting to explore if the $\Theta$-term has any physical effect in the early universe.

In the context of the AdS/CFT correspondence, the gravitational $\Theta$-term is dual to field theories with  non-vanishing contact terms of the energy-momentum tensor. As a consequence, in the presence of the $\Theta$-term the $(2+1)-$dimensional dual gauge theory acquires the same third order parity violating transport coefficient $\vartheta$. We have studied various properties of this new transport coefficient $\vartheta$. Historically, gauge/gravity duality has played a significant role in hydrodynamics by discovering new universal effects \cite{Bhattacharyya:2008jc,Erdmenger:2008rm,Son:2009tf}. This is another nice example where gauge/gravity duality teaches us about a new hydrodynamic effect. However, we would like to note that one could have found this hydrodynamic effect even without knowing anything about the AdS/CFT correspondence, simply by studying the effect of the $\Theta$-term on the stretched horizon. 

It is important to note that our conclusion about the effect of the $\Theta$-term on the stretched horizon depends only on the near horizon geometry but not on the details of the metric and hence it can easily be generalized for arbitrary cosmological horizons. The AdS/CFT correspondence has taught us that the membrane paradigm fluid on the black hole horizon and   linear response of a strongly coupled quantum field theory in the low frequency limit are related. However, it is not at all clear if this connection between the membrane paradigm and holography goes beyond the AdS/CFT correspondence. In particular, it will be extremely interesting to figure out if the same conclusion is true for holographic models of cosmological spacetime. 

\section*{Acknowledgments}
We would like to thank T. Hartman, T. Jacobson, S. Jain and M. Rangamani  for the useful discussions. The work of WF was supported by the National Science Foundation under Grant Numbers PHY-1316033 and by the Texas Cosmology Center, which is supported by the College of Natural Sciences and the Department of Astronomy at the University of Texas at Austin and the McDonald Observatory. The work of SK was supported by NSF grant PHY-1316222.
\appendix

\section{Near horizon metric}\label{nh}
We will denote the 3-dimensional absolute event horizon by $\H$. We have defined a well-behaved time coordinate $\t$ on the horizon as well as spatial coordinates $x^A$ with $A=1,2$. Therefore, the event horizon $\H$ has coordinates $x^i\equiv \left(\t, x^A\right)$ with a metric $h_{ij}$. We can always choose coordinates $x^i\equiv \left(\t, x^A\right)$ such that null generator \begin{equation}
l=\frac{\partial}{\partial \t}\ .
\end{equation}
Where we have chosen the spatial coordinates $x^A$ such that they comove with the horizon generator $l$ and $h_{0A}=0$. The basis $e_i\equiv (l, e_A)$ spans the horizon and $l.e_A=0$. The induced spatial metric on a constant $\t$ hypersurface is $\gamma_{AB}$.  Following \cite{Price:1986yy}, let us now introduce a future directed ingoing null vector $k^\mu$ at each point on $\H$ which obeys
\begin{equation}\label{keqn}
l.k=-1\ , \qquad k.e_A=0\ .
\end{equation} 
There exists a unique congruence of null ingoing geodesics that are tangent to $k^\mu$ on the horizon. Horizon coordinates $x^i\equiv \left(\t, x^A\right)$ can be carried on these geodesics into the near horizon region. We can use the affine parameter $\lambda$ on the null geodesics as the fourth coordinate, where on the horizon $\H$
\begin{equation}
\lambda=0 \qquad \text{and} \qquad k=-\frac{\partial}{\partial \lambda}\ .
\end{equation}
We will use these coordinates $(\lambda, \t, x^A)$ to explore the near horizon region of a black hole. Since $k^\mu$ is tangent to affinely parametrized null geodesic, it obeys
\begin{equation}
k^\mu \nabla_\mu k^\nu=0
\end{equation}
and this equation leads to
\begin{equation}
\partial_\lambda g_{\mu \lambda}=0 .
\end{equation}
Therefore, $g_{\mu \lambda}$ are $\lambda -$independent and from equation (\ref{keqn}) in this coordinate system we obtain
\begin{equation}
g_{\lambda \lambda}=0 \ , \qquad g_{\lambda A}=0\ , \qquad g_{\t \lambda}=1\ .
\end{equation}
On the horizon $\H$: $g_{\t \t}=g_{\t A}=0$. From the equation (\ref{gh}) one can easily show that
\begin{equation}
\frac{1}{2}\partial_\lambda g_{\t \t}=-g_H
\end{equation}
and hence near the horizon 
\begin{equation}
g_{\t  \t}=-2 g_H \lambda +\O (\lambda^2)\ ,
\end{equation}
where, $g_H$ is the surface gravity defined by equation (\ref{gh}). Similarly, one can write
\begin{equation}
g_{\t A}=-2\Omega_A(\t, x) \lambda +\O (\lambda^2)
\end{equation}
where $\Omega_A(\t, x)$ is the Hajicek field defined in the following way
\begin{equation}
\Omega_A=\langle d\t, \nabla_A l \rangle\ .
\end{equation}
This quantity is related to the angular momentum of a black hole. Therefore, near the horizon space-time metric has the following form
\begin{align}
ds^2=-2 g_H(\t,x) \lambda~d\t^2+ 2d\t d\lambda+\gamma_{AB}(\lambda,\t,x)\left(dx^A-2\Omega^A(\t, x) \lambda d\t \right)&\left(dx^B-2\Omega^B(\t, x) \lambda d\t \right)\nonumber\\
&+\O (\lambda^2).
\end{align}
We will now restrict to the simpler case: $g_H(\t,x) =$constant on the horizon $\H$.  For the case of constant $g_H$ (and non-zero), we can define a new radial coordinate $2 g_H \lambda =r^2$ and the near horizon geometry has the form 
\begin{equation}
ds^2=-r^2~d\t^2+ \frac{2r}{g_H}d\t dr+\gamma_{AB}(r,\t,x)\left(dx^A-\frac{\Omega^A(\t, x)}{g_H} r^2 d\t \right)\left(dx^B-\frac{\Omega^B(\t, x)}{g_H} r^2 d\t \right)+\O (r^4)
\end{equation}
where the horizon is at $r=0$.
\section{Stretched horizon: transport coefficients}\label{transport}

The fluid living on the stretched horizon is almost at rest in the comoving coordinates, i.e., 
\be
u^\t=\frac{1}{\epsilon}\ , \qquad u^A= \O(\epsilon)\ .
\ee
Therefore, $u^i= U^i +\O(\epsilon)$, where $U^\mu$ is the velocity of the FIDOs. We can now easily show that in the limit $\epsilon \rightarrow 0$
\be
D^{(3)}_i u^i=\frac{1}{\epsilon}\theta_H
\ee
and
\be
f_{00}=f_{0A}=\O(\epsilon)\ , \qquad f_{AB}=\frac{2}{\epsilon}\sigma^H_{AB}+\O(\epsilon)\ .
\ee

Let us now investigate the energy momentum tensor on the stretched horizon (\ref{membrane_em}). We can rewrite equation (\ref{membrane_em}) in the following form
\begin{align}
\T^{i j}=&\m^2\left[\left(-K-K_{kl}u^k u^l \right)u^i u^j+\frac{1}{2}\left(K-K_{kl}u^k u^l \right)P^{ij} \right]_\M\nonumber \\
&+ \m^2\left[ \frac{1}{2}K P^{ij}-K^{ij}+ \left(K_{kl}u^k u^l\right)\left(u^i u^j+\frac{1}{2}P^{ij}\right)\right]_\M .
\end{align}
Therefore, comparing equation (\ref{newton}) with the last equation, we obtain,
\begin{align}
\rho&=\m^2\left[-K-K_{kl}u^k u^l \right]_\M \ ,\label{rho}\\
p- \zeta D^{(3)}_m u^m&=\m^2\left[\frac{1}{2}\left(K-K_{kl}u^k u^l \right) \right]_\M\ ,\label{p}\\
 -\eta P^{ik}P^{jl}  f_{kl} &=\m^2\left[ \frac{1}{2}K P^{ij}-K^{ij}+ \left(K_{kl}u^k u^l\right)\left(u^i u^j+\frac{1}{2}P^{ij}\right)\right]_\M\ .\label{eta}
\end{align}
From equations (\ref{rho}) and (\ref{p}), we can find out $\rho$, $p$ and $\zeta$ easily. For $\eta$ one needs to look at the spatial components (i.e. $(...)^{AB}$ components) of equations (\ref{eta}). So finally we obtain \cite{Price:1986yy}
\begin{align}
&\rho=-\frac{\m^2}{\epsilon} \theta_H\ ,\\
& p=\frac{\m^2}{\epsilon} g_H\ ,\\
& \zeta=- \frac{\m^2}{2} \ ,\\
&\eta=\frac{\m^2}{2} \ .
\end{align}
We provide an alternative derivation of these relations in section \ref{potot}.


\end{document}